\documentclass{ws-mpla}

\begin{document}

\markboth{}
{}

%%%%%%%%%%%%%%%%%%%%% Publisher's Area please ignore %%%%%%%%%%%%%%
\catchline{}{}{}{}{}
%%%%%%%%%%%%%%%%%%%%%%%%%%%%%%%%%%%%%%%%%%%%%%%%%%%%%%%%%%%%%%%%%%%

\title{$l_{\alpha}\rightarrow 3l_{\beta}$ in Minimal R-symmetric Supersymmetric Standard Model}

\author{Ke-Sheng Sun    $^{a,\ast}$,
        Jian-Bin Chen   $^{b,\star}$,
        Hai-Bin Zhang   $^{c,\dagger}$,
        Sheng-Kai Cui   $^{c,\ddagger}$}

\address{$^{a}$   Department of Physics, Baoding University, Baoding 071000,China\\
         $^{b}$   College of Physics and Optoelectronic Engineering, Taiyuan University of Technology, Taiyuan 030024, China\\
         $^{c}$   Department of Physics, Hebei University, Baoding 071002, China\\
         $^{\ast}   $    sunkesheng@126.com;sunkesheng@mail.dlut.edu.cn\\
         $^{\star}  $    chenjianbin@tyut.edu.cn\\
         $^{\dagger}$    hbzhang@hbu.edu.cn\\
         $^{\ddagger}$   2252953633@qq.com}

\maketitle

\pub{Received (Day Month Year)}{Revised (Day Month Year)}

\begin{abstract}
Lepton flavor violation decays are channels which may lead to fundamental discoveries in the
forthcoming years and this make it an exciting research field for beyond the Standard Model
searches. In this work, we present an analysis of the lepton flavor violation decays $l_{\alpha}\rightarrow 3l_{\beta}$ in Minimal R-symmetric Supersymmetric Standard Model. The prediction for BR($l_{\alpha}\rightarrow 3l_{\beta}$) depend on the off-diagonal entries of the slepton mass matrix. The contributions to Wilson coefficients can be classified into Higgs penguins, photon penguins, Z penguins, and box diagrams. It shows the contribution from Z penguins dominates the predictions for BR($l_{\alpha}\rightarrow 3l_{\beta}$), and the contributions from Higgs penguins and box diagrams play different roles in different decay channels. The theoretical predictions for BR($l_{\alpha}\rightarrow 3l_{\beta}$) can reach the future experimental limits, and there channels are very promising to be observed in near future experiment.

\keywords{R-symmetry; MRSSM; Lepton flavor violation}
\end{abstract}

\section{Introduction}\label{intro}	

Many efforts have been devoted to searching for Lepton Flavor Violation (LFV) decays in experiment and literature, since it is one of the signals for New Physics (NP) beyond the Standard Model (SM) in which the lepton flavor is conserved.
The present upper bounds and future sensitivities for the LFV decays $l_{\alpha}\rightarrow 3l_{\beta}$ are summarized in Table.\ref{Tab1}. Several predictions for these LFV processes have obtained in the framework of various extended SM. One of the most attractive concepts for NP beyond SM is supersymmetry, which is the only possible nontrivial extension of the Poincar$\acute{\text{e}}$ algebra in a relativistic quantum field theory.

In this work, we will analyze these LFV decays in the Minimal R-symmetric Supersymmetric Standard Model (MRSSM). The MRSSM is proposed in Ref.\cite{Kribs} and gives a new solution to the supersymmetric flavor problem in MSSM, where the R-symmetry, being different from R-parity, is a fundamental symmetry proposed several years ago\cite{Fayet,Salam} and not present in models like the Minimal Supersymmetric Standard Models(MSSM). The continuous R-symmetry forbids Majorana gaugino masses, then the gaugino masses can not be anything but Dirac masses which leads to the gauge boson has a Dirac gaugino and a scalar superpartner. The R-symmetry also forbids $\mu$ term, A terms, and all left-right squark and slepton mass mixings. The $R$-charged Higgs $SU(2)_L$ doublets $\hat{R}_u$ and $\hat{R}_d$ are introduced in MRSSM to yield the Dirac mass terms of higgsinos. Additional superfields $\hat{S}$, $\hat{T}$ and $\hat{O}$ are introduced to yield Dirac mass terms of gauginos. Studies on phenomenology in MRSSM can be found in literatures \cite{Die1,Die2,Die3,Die4,Die5,Die6,Kumar,Blechman,Kribs1,Frugiuele,Jan,Chakraborty,Braathen,Athron,Alvarado,sks1,sks2,kss}.

\begin{table}[h]
\tbl{Present limits and future sensitivities for BR($l_{\alpha}\rightarrow 3l_{\beta}$).}
{\begin{tabular}{@{}ccc@{}}
\toprule
LFV process& Present limit& Future sensitivity\\
\colrule
$\mu\rightarrow 3e$&$1.0\times 10^{-12}$ Ref.\cite{tb1}&$ 10^{-16}$ Ref.\cite{tb2}\\
$\tau\rightarrow 3e$&$2.7\times 10^{-8}$ Ref.\cite{tb4}&$10^{-9}-10^{-10}$ Ref.\cite{tb3}\\
$\tau\rightarrow 3\mu$&$2.1\times 10^{-8}$ Ref.\cite{tb4}&$10^{-9}-10^{-10}$ Ref.\cite{tb3}\\
\botrule
\end{tabular}\label{Tab1}}
\end{table}

In SM, the LFV decays mainly originate from the charged current with the mixing among three lepton generations.
The fields of the flavor neutrinos in charged current weak interaction Lagrangian are combinations of three massive neutrinos:
\begin{eqnarray}
{\cal L}& = &-\frac{g_2}{\sqrt{2}}\sum_{l=e,\mu,\tau}\overline{l_{L}}(x)\gamma_\mu \nu_{lL}(x)W^\mu(x)+h.c.,
\nonumber\\
\nu_{lL}(x)& = &\sum_{i=1}^{3}\Big(U_{PMNS}\Big)_{li}\nu_{iL}(x),\nonumber
\end{eqnarray}
where $g_2$ denotes the coupling constant of gauge group SU(2), $\nu_{lL}$ are fields of the flavor neutrinos, $\nu_{iL}$ are fields of massive neutrinos, and $U_{PMNS}$ corresponds to the unitary neutrino mixing matrix \cite{Pontecorvo1,Pontecorvo2,Maki}.

In this paper, we have studied the LFV decays $l_{\alpha}\rightarrow 3l_{\beta}$ in MRSSM by considering the constraints on off-diagonal entires $\delta^{ij}$ from LFV decays $l_{\alpha}\rightarrow l_{\beta}\gamma$. We first consider an effective Lagrangian that includes the operators relevant for the flavor observable of $l_{\alpha}\rightarrow 3l_{\beta}$. Then, by taking into account all possible 1-loop topologies leading to the relevant operators, the Wilson coefficients are computed for each Feynman diagram, in which the contributions have been classified into four categories (Higgs, photon, Z, box). Finally, the results for the Wilson coefficients are plugged in a general expression for BR($l_{\alpha}\rightarrow 3l_{\beta}$) and a final result is obtained.

The paper is organized as follows. In Section \ref{sec2}, we firstly provide a brief introduction on MRSSM. Then, we derive the analytic expressions of the Wilson coefficients in each Feynman diagram contributing to $l_{\alpha}\rightarrow 3l_{\beta}$ in MRSSM in detail. The numerical results are presented in Section \ref{sec3}, and the conclusion is drawn in Section \ref{sec4}.
\section{MRSSM}
\label{sec2}
First, it is necessary to provide a simple introduction to MRSSM. In MRSSM, the spectrum of fields contain the standard MSSM matter, Higgs and gauge superfields augmented by chiral adjoints, two R-Higgs
iso-doublets. The superfields with R-charge in MRSSM can be found in Ref.\cite{sks2}, which is not listed for simplicity. The general form of the superpotential in MRSSM is given by\cite{Die1}
\begin{eqnarray}
{\cal W}_{MRSSM} &=& \mu_d(\hat{R}_dH_d)+\mu_u(\hat{R}_uH_u)+\Lambda_d(\hat{R}_d\hat{T})H_d\nonumber\\
&+&\Lambda_u(\hat{R}_u\hat{T})H_u+Y_u\bar{U}(QH_u)-Y_d\bar{D}(QH_d)\nonumber\\
&-&Y_e\bar{E}(LH_d)+\lambda_d\hat{S}(\hat{R}_dH_d)+\lambda_u\hat{S}(\hat{R}_uH_u),
\end{eqnarray}
where $H_u$ and $H_d$ stand for the MSSM-like Higgs weak iso-doublets, $\hat{R}_u$ and $\hat{R}_d$ stand for the $R$-charged Higgs $SU(2)_L$ doublets and the corresponding Dirac higgsino mass parameters are $\mu_u$ and $\mu_d$.
The Yukawa-like trilinear terms, which involve the singlet $\hat{S}$ and the triplet $\hat{T}$, contain four parameters $\lambda_u$, $\lambda_d$, $\Lambda_u$ and $\Lambda_d$. The triplet $\hat{T}$ is given by
\begin{equation}
\hat{T} = \left(
\begin{array}{cc}
\hat{T}^0/\sqrt{2} &\hat{T}^+ \\
\hat{T}^-  &-\hat{T}^0/\sqrt{2}\end{array}
\right).
 \end{equation}
The soft-breaking scalar mass terms are given by
\begin{eqnarray}
V_{SB,S} &=& m^2_{H_d}(|H^0_d|^2+|H^{-}_d|^2)+m^2_{H_u}(|H^0_u|^2+|H^{+}_u|^2)+m^2_{R_u}(|R^0_u|^2+|R^{-}_u|^2)\nonumber\\
&+&m^2_{R_d}(|R^0_d|^2+|R^{+}_d|^2)+(B_{\mu}(H^-_dH^+_u-H^0_dH^0_u)+h.c.)\nonumber\\
&+&\tilde{d}^*_{L,i} m_{q,{i j}}^{2} \tilde{d}_{L,j} +\tilde{d}^*_{R,i} m_{d,{i j}}^{2} \tilde{d}_{R,j}+\tilde{u}^*_{L,i}  m_{q,{i j}}^{2} \tilde{u}_{L,j}+\tilde{u}^*_{R,i}  m_{u,{i j}}^{2} \tilde{u}_{R,j}\nonumber\\
&+&\tilde{e}^*_{L,i} m_{l,{i j}}^{2} \tilde{e}_{L,j}+\tilde{e}^*_{R,{i}} m_{r,{i j}}^{2} \tilde{e}_{R,{j}} +\tilde{\nu}^*_{L,i} m_{l,{i j}}^{2} \tilde{\nu}_{L,j}\nonumber\\
&+&m^2_S|S|^2+ m^2_O|O^2|+m^2_T(|T^0|^2+|T^-|^2+|T^+|^2).
\end{eqnarray}
It is noted worthwhile that all trilinear scalar couplings involving Higgs bosons to squarks and sleptons are forbidden due to the $R$-symmetry. The soft-breaking Dirac mass terms of the singlet $\hat{S}$, triplet $\hat{T}$ and octet $\hat{O}$ take the form
\begin{equation}
V_{SB,DG}=M^B_D\tilde{B}\tilde{S}+M^W_D\tilde{W}^a\tilde{T}^a+M^O_D\tilde{g}\tilde{O}+h.c.,
\label{}
\end{equation}
where $\tilde{B}$, $\tilde{W}$ and $\tilde{g}$ are usually MSSM Weyl fermions.

For convenience, we will use the notations in Ref.\cite{sks1,sks2} for the mass matrices and mixing matrices of neutralino, chargino, slepton and sneutrino. One can find the explicit expressions of these mass matrices and mixing matrices in Ref.\cite{sks1,sks2} and we will not listed them in following. In the basis $(\sigma_d,\sigma_u,\sigma_S,\sigma_T)$, the pseudo-scalar Higgs boson mass matrix takes a simple form
\begin{eqnarray}
{\cal M}^2_{A^0} &=& \left(
\begin{array}{cccc}
B_{\mu} \frac{v_u}{v_d} & B_{\mu} & 0 & 0 \\
B_{\mu} &  B_{\mu} \frac{v_d}{v_u} & 0 & 0 \\
0 & 0 & m_S^2+\frac{\lambda_d^2 v_d^2+\lambda_u^2 v_u^2 }{2} & \frac{\lambda_d\Lambda_d v_d^2-\lambda_u\Lambda_u v_u^2}{2 \sqrt{2}} \\
0 & 0 & \frac{\lambda_d\Lambda_d v_d^2-\lambda_u\Lambda_u v_u^2}{2 \sqrt{2}}& m_T^2+ \frac{\Lambda_d^2 v_d^2+\Lambda_u^2 v_u^2 }{4}\\
\end{array}
\right),
\end{eqnarray}
and is diagonalized by unitary matrix $Z^{A}$
\begin{equation}
Z^A {\cal M}^2_{A^0} (Z^{A})^{\dagger}.
\end{equation}
In the weak basis $(\phi_d,\phi_u,\phi_S,\phi_T)$, the scalar Higgs boson mass matrix is given by
\begin{eqnarray}
{\cal M}^2_h &=& \left(
\begin{array}{cc}
{\cal M}_{11}&{\cal M}_{21}^T\\
{\cal M}_{21}&{\cal M}_{22}\\
\end{array}
\right),
\end{eqnarray}
where the submatrices ($c_{\beta}=cos\beta$, $s_{\beta}=sin\beta$) are
\begin{eqnarray}
{\cal M}_{11}&=& \left(
\begin{array}{cc}
m_Z^2c^2_{\beta}+m_A^2s^2_{\beta}&-(m_Z^2+m_A^2)s_{\beta}c_{\beta}\\
-(m_Z^2+m_A^2)s_{\beta}c_{\beta}&m_Z^2s^2_{\beta}+m_A^2c^2_{\beta}\\
\end{array}
\right),\nonumber\\
{\cal M}_{21}&=& \left(
\begin{array}{cc}
v_d(\sqrt{2}\lambda_d\mu_d^{eff,+}-g_1M_B^D)&
v_u(\sqrt{2}\lambda_u\mu_u^{eff,-}+g_1M_B^D) \\
v_d(\Lambda_d\mu_d^{eff,+}+g_2M_W^D)& -
v_u(\Lambda_u\mu_u^{eff,1}+g_2M_W^D) \\
\end{array}
\right),\nonumber\\
{\cal M}_{22}&=& \left(
\begin{array}{cc}
4 (M_B^D)^2+m_S^2+\frac{\lambda_d^2 v_d^2+\lambda_u^2 v_u^2}{2} \;
& \frac{\lambda_d \Lambda_d v_d^2-\lambda_u \Lambda_u v_u^2}{2 \sqrt{2}} \\
 \frac{\lambda_d \Lambda_d v_d^2-\lambda_u \Lambda_u v_u^2}{2 \sqrt{2}} \;
 & 4 (M_W^D)^2+m_T^2+\frac{\Lambda_d^2 v_d^2+\Lambda_u^2 v_u^2}{4}\\
\end{array}
\right),\nonumber
\end{eqnarray}
and is diagonalized by unitary matrix $Z^{h}$
\begin{equation}
Z^h {\cal M}^2_{h} (Z^{h})^{\dagger}.
\end{equation}
The modified $\mu_i$ parameters are given by
\begin{align}
\mu_d^{eff,+}&= \frac{1}{2} \Lambda_d v_T  + \frac{1}{\sqrt{2}} \lambda_d v_S  + \mu_d ,\nonumber\\
\mu_u^{eff,-}&= -\frac{1}{2} \Lambda_u v_T  + \frac{1}{\sqrt{2}} \lambda_u v_S  + \mu_u.\nonumber
\end{align}
The $v_T$ and $v_S$ are vacuum expectation values of $\hat{T}$ and $\hat{S}$ which carry zero $R$-charge.

The relevant Lagrangian for $l_{\alpha}\rightarrow 3l_{\beta}$ can be written as \cite{Flavor}
\begin{equation}
{\cal L}_{LFV}={\cal L}_{ll\gamma}+{\cal L}_{4l}.\\
\end{equation}
The $ll\gamma$ interaction is given by
\begin{equation}
{\cal L}_{ll\gamma}=e\bar{l}_{\beta}[\gamma^\mu(K^L_1 P_L+K^R_1 P_R)+i m_{l_{\alpha}}\sigma^{\mu\nu}q_{\nu}(K^L_2 P_L+K^R_2 P_R)]l_{\alpha}A_{\mu}+h.c..
\label{llg}
\end{equation}
The general $4l$ 4-fermion interaction Lagrangian can be written as
\begin{equation}
{\cal L}_{4l}=A^{I}_{XY}\bar{l}_{\beta}\Gamma_{I}P_Xl_{\alpha}\bar{l}_{\beta}\Gamma_{I}P_Yl_{\beta}+h.c.,
\label{4l}
\end{equation}
where $I=\{S,V,T\}$, $X,Y=\{L,R\}$, $\Gamma_S=1$, $\Gamma_V=\gamma_{\mu}$ and $\Gamma_T=\sigma_{\mu\nu}$.

%%%%%%%%%%%%%%%%%%%%%%%%%%%%%%%%%%%%%%%%%%%%%%%%%%%%%%%%%%%%%%%%%%%
\begin{figure}[htbp]
\setlength{\unitlength}{1mm}
\begin{minipage}[c]{1\textwidth}
\centering
\includegraphics[width=4.0in]{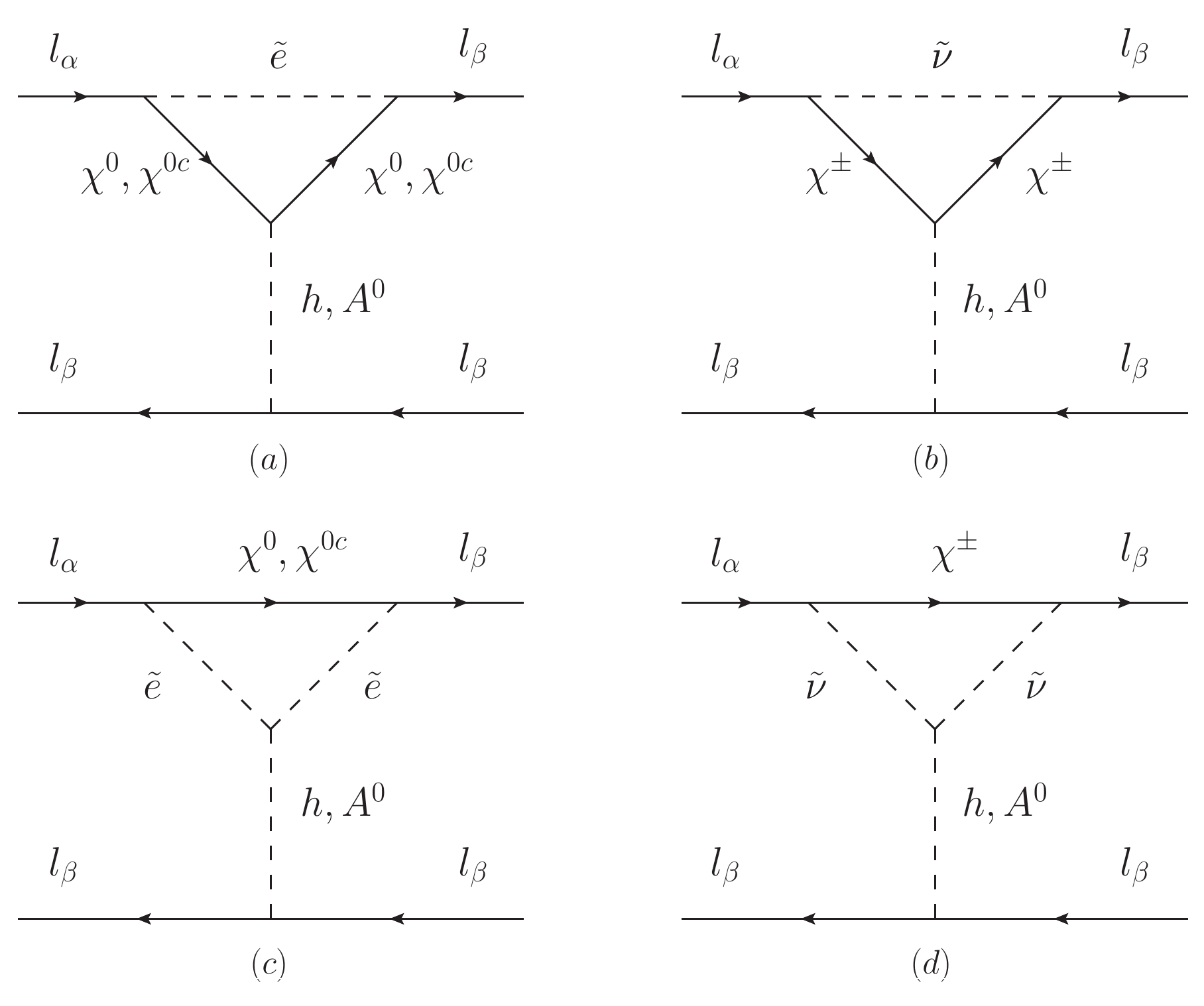}
\end{minipage}
\caption[]{Higgs penguin diagrams contributing to $l_{\alpha}\rightarrow 3l_{\beta}$ in MRSSM.}
\label{h}
\end{figure}
%%%%%%%%%%%%%%%%%%%%%%%%%%%%%%%%%%%%%%%%%%%%%%%%%%%%%%%%%%%%%%%%%%%

The Higgs mediated diagrams contributing to $l_{\alpha}\rightarrow 3l_{\beta}$ in MRSSM are presented in Fig.\ref{h}. The coefficients in Fig.\ref{h} (a,b) are calculated by
\begin{eqnarray}
A^S_{XY}&&=\frac{-1}{M_H^2}C^1_XC^4_Y\big(C^2_{X'}C^3_X{\cal B}_0(0,M_2,M_1)+(C^2_XC^3_XM_1M_2+C^2_XC^3_{X'}M_1m_{l_{\alpha}}\nonumber\\
&&+C^2_{X'}C^3_XM_3^2){\cal C}_0+m_{l_{\alpha}}(C^2_XC^3_{X'}M_1+C^2_{X'}C^3_Xm_{l_{\alpha}}+C^2_{X'}C^3_{X'}M_2){\cal C}_1\big),
\end{eqnarray}
where $M_H$ denote $m_h$ or $m_{A^0}$. The symbols $M_1$, $M_2$ and $M_3$ denote masses of sparticles in internal lines. The symbols $X'(Y')$ are defined as
\begin{displaymath}
X'(Y') =
\left\{ \begin{array}{ll}
L,  & \text{when}\;\; X(Y)=R,\\
R, &  \text{when}\;\; X(Y)=L.
\end{array} \right.
\end{displaymath}
Here and following, ${\cal B}$, ${\cal C}_0$ and ${\cal C}_1$ denote the Passarino-Veltman integrals, where the masses of outgoing leptons are set as zero. The explicit expressions of these intergrals will be introduced later on.
The couplings $C^4_X$ are identical in Fig.\ref{h}(a-d),
\begin{eqnarray}
C^4_L&&=C^4_R=-\frac{i}{\sqrt{2}}Y_{l_{\beta}}Z^h_{l1},\text{ $h$ mediated diagrams},\nonumber\\
C^4_L&&=-C^4_R=\frac{1}{\sqrt{2}}Y_{l_{\beta}}Z^A_{l1},\text{ $A^0$ mediated diagrams},
\end{eqnarray}
however other couplings are defined different for each diagram. For $h$ and $\chi^0$ mediated diagram in Fig.\ref{h}(a), the relevant couplings and masses denotation are
\begin{eqnarray}
C^1_L&&=-i\sqrt{2}N^{1\ast}_{i1}Z^{E\ast}_{k(3+\beta)},C^1_R=-iY_{l_{\beta}}Z^{E\ast}_{k(3+\beta)}N^2_{i3},\nonumber\\
C^2_L&&=\frac{i}{2}\big(-g_2N^{1\ast}_{j2}N^{2\ast}_{i3}Z^h_{l1}-\sqrt{2}\lambda_uN^{1\ast}_{j4}N^{2\ast}_{i1}Z^h_{l2}+
\Lambda_uN^{1\ast}_{j4}N^{2\ast}_{i2}Z^h_{l2}+g_2N^{1\ast}_{j2}N^{2\ast}_{i4}Z^h_{l2}\nonumber\\
&&+g_1N^{1\ast}_{j1}(N^{2\ast}_{i3}Z^h_{l1}-N^{2\ast}_{i4}Z^h_{l2})-\sqrt{2}
\lambda_uN^{1\ast}_{j4}N^{2\ast}_{i4}Z^h_{l3}+\Lambda_uN^{1\ast}_{j4}N^{2\ast}_{i4}Z^h_{l4}
\nonumber\\
&&+N^{1\ast}_{j3}(\Lambda_dN^{2\ast}_{i2}Z^h_{l1}+N^{2\ast}_{i3}(\Lambda_dZ^h_{l4}
+\sqrt{2}\lambda_dZ^h_{l3})+\sqrt{2}\lambda_dN^{2\ast}_{i1}Z^h_{l1})\big),\nonumber\\
C^2_R&&=\frac{i}{2}\big(\Lambda_dZ^h_{l1}N^{1}_{i3}N^{2}_{j2}+\Lambda_uZ^h_{l2}
N^{1}_{i4}N^{2}_{j2}+g_1Z^h_{l1}N^1_{i1}N^2_{j3}-g_2Z^h_{l2}N^1_{i2}N^2_{j3}
\nonumber\\
&&+\Lambda_dZ^h_{l4}N^{1}_{i3}N^{2}_{j3}+\sqrt{2}\Lambda_d N^{1}_{i3}(Z^h_{l1}N^{2}_{j1}+Z^h_{l3}N^{2}_{j3})-g_1Z^h_{l2}N^1_{i1}N^2_{j4}
\nonumber\\
&&+g_2Z^h_{l2}N^1_{i2}N^2_{j4}+\Lambda_uZ^h_{l4}N^{1}_{i4}N^{2}_{j4}-\sqrt{2}\Lambda_u N^{1}_{i4}(Z^h_{l2}N^{2}_{j1}+Z^h_{l3}N^{2}_{j4})\big),\nonumber\\
C^3_L&&=-iN^{2\ast}_{j3}Y_{l_{\alpha}}Z^E_{k(3+\alpha)},C^3_R=-i\sqrt{2}g_1Z^E_{k(3+\alpha)}N^1_{j1},\nonumber\\
M_1&&=m^i_{\chi^0},M_2=m^j_{\chi^0},M_3=m^k_{\tilde{e}}.\label{E1}
\end{eqnarray}
For $A^0$ and $\chi^0$ mediated diagram in Fig.\ref{h}(a), the couplings $C^1_X$, $C^3_X$ and masses denotation are same with those in Eq.(\ref{E1}), the other couplings are
\begin{eqnarray}
C^2_L&&=\frac{1}{2}\big(-g_2N^{1\ast}_{j2}N^{2\ast}_{i3}Z^A_{l1}+\sqrt{2}\lambda_uN^{1\ast}_{j4}N^{2\ast}_{i1}Z^A_{l2}-
\Lambda_uN^{1\ast}_{j4}N^{2\ast}_{i2}Z^A_{l2}+g_2N^{1\ast}_{j2}N^{2\ast}_{i4}Z^A_{l2}\nonumber\\
&&+g_1N^{1\ast}_{j1}(N^{2\ast}_{i4}Z^A_{l2}-N^{2\ast}_{i2}Z^A_{l1})+\sqrt{2}
\lambda_uN^{1\ast}_{j4}N^{2\ast}_{i4}Z^A_{l3}-\Lambda_uN^{1\ast}_{j4}N^{2\ast}_{i4}Z^A_{l4}
\nonumber\\
&&-N^{1\ast}_{j3}(\Lambda_dN^{2\ast}_{i2}Z^A_{l1}+N^{2\ast}_{i3}(\Lambda_dZ^A_{l4}
+\sqrt{2}\lambda_dZ^A_{l3})+\sqrt{2}\lambda_dN^{2\ast}_{i1}Z^A_{l1})\big),\nonumber\\
C^2_R&&=\frac{1}{2}(\Lambda_dZ^A_{l1}N^{1}_{i3}N^{2}_{j2}+\Lambda_uZ^A_{l2}
N^{1}_{i4}N^{2}_{j2}-g_1Z^A_{l1}N^1_{i1}N^2_{j3}+g_2Z^A_{l2}N^1_{i2}N^2_{j3}
\nonumber\\
&&+\Lambda_dZ^A_{l4}N^{1}_{i3}N^{2}_{j3}+\sqrt{2}\Lambda_d N^{1}_{i3}(Z^A_{l1}N^{2}_{j1}+Z^A_{l3}N^{2}_{j3})+g_1Z^A_{l2}N^1_{i1}N^2_{j4}
\nonumber\\
&&-g_2Z^A_{l2}N^1_{i2}N^2_{j4}+\Lambda_uZ^A_{l4}N^{1}_{i4}N^{2}_{j4}-\sqrt{2}\Lambda_u N^{1}_{i4}(Z^A_{l2}N^{2}_{j1}+Z^A_{l3}N^{2}_{j4})\big).\label{E2}
\end{eqnarray}
For $h$ and $\chi^{0c}$ mediated diagram in Fig.\ref{h}(a), the couplings $C^2_X$ are same with those in Eq.(\ref{E1}), the other couplings and masses denotation are
\begin{eqnarray}
C^1_L&&=-iN^{2\ast}_{i3}Z^E_{k\beta}Y_{l_{\beta}},
C^1_R=\frac{i}{\sqrt{2}}Z^{E\ast}_{k\beta}(g_1N^1_{i1}+g_2N^1_{i2}),\nonumber\\
C^3_L&&=\frac{i}{\sqrt{2}}Z^{E}_{k\alpha}(g_1N^{1\ast}_{j1}+g_2N^{1\ast}_{j2}),
C^3_R=-iY_{l_{\alpha}}Z^{E}_{k\alpha}N^{2\ast}_{j3},\nonumber\\
M_1&&=m^i_{\chi^{0c}},M_2=m^j_{\chi^{0c}},M_3=m^k_{\tilde{e}}.\label{E3}
\end{eqnarray}
For $A^0$ and $\chi^{0c}$ mediated diagram in Fig.\ref{h}(a), the couplings $C^2_X$ are same with those in Eq.(\ref{E2}), couplings $C^1_X$, $C^3_X$ and masses denotation are same with those in Eq.(\ref{E3}).

For $h$ and $\chi^{\pm}$ mediated diagram in Fig.\ref{h}(b), the relevant couplings and masses denotation are
\begin{eqnarray}
C^1_L&&=iU^{1\ast}_{i2}Z^V_{k\beta}Y_{l_{\beta}},
C^1_R=-ig_2Z^{V\ast}_{k\beta}V^1_{i1},\nonumber\\
C^2_L&&=\frac{-i}{2}\big(U^{1\ast}_{i1}(2g_2V^{1\ast}_{j1}Z^h_{l4}+\sqrt{2}\Lambda_dV^{1\ast}_{j2}Z^h_{l1})
\nonumber\\
&&+U^{1\ast}_{i2}(\sqrt{2}g_2V^{1\ast}_{j1}Z^h_{l1}+\sqrt{2}\lambda_dV^{1\ast}_{j2}Z^h_{l3}
-\Lambda_dV^{1\ast}_{j2}Z^h_{l4})\big),\nonumber\\
C^2_R&&=\frac{-i}{2}\big(U^{1}_{j1}(2g_2V^{1}_{i1}Z^h_{l4}+\sqrt{2}\Lambda_dV^{1}_{i2}Z^h_{l1})
\nonumber\\
&&+U^{1\ast}_{j2}(\sqrt{2}g_2V^{1}_{i1}Z^h_{l1}+\sqrt{2}\lambda_dV^{1}_{i2}Z^h_{l3}
-\Lambda_dV^{1}_{i2}Z^h_{l4})\big),\nonumber\\
C^3_L&&=-ig_2V^{1\ast}_{j1}Z^V_{k\alpha},
C^3_R=iY_{l_{\alpha}}Z^{V}_{k\alpha}U^{1}_{j2},\nonumber\\
M_1&&=m^i_{\chi^{\pm}},M_2=m^j_{\chi^{\pm}},M_3=m^k_{\tilde{\nu}}.\label{E4}
\end{eqnarray}
For $A^0$ and $\chi^{\pm}$ mediated diagram in Fig.\ref{h}(b), the couplings $C^1_X$, $C^3_X$ and masses denotation are same with those in Eq.(\ref{E4}), and the remaining couplings are
\begin{eqnarray}
C^2_L&&=\frac{-1}{2}\big(U^{1\ast}_{i1}(2g_2V^{1\ast}_{j1}Z^A_{l4}+\sqrt{2}\Lambda_dV^{1\ast}_{j2}Z^A_{l1})
\nonumber\\
&&+U^{1\ast}_{i2}(\sqrt{2}g_2V^{1\ast}_{j1}Z^A_{l1}-\sqrt{2}\lambda_dV^{1\ast}_{j2}Z^A_{l3}
+\Lambda_dV^{1\ast}_{j2}Z^A_{l4})\big),\nonumber\\
C^2_R&&=\frac{1}{2}\big(U^{1}_{j1}(2g_2V^{1}_{i1}Z^A_{l4}-\sqrt{2}\Lambda_dV^{1}_{i2}Z^A_{l1})
\nonumber\\
&&+U^{1\ast}_{j2}(\sqrt{2}g_2V^{1}_{i1}Z^A_{l1}-\sqrt{2}\lambda_dV^{1}_{i2}Z^A_{l3}
+\Lambda_dV^{1}_{i2}Z^A_{l4})\big).\label{E5}
\end{eqnarray}

The coefficients in Fig.\ref{h} (c,d) are calculated by
\begin{eqnarray}
A^S_{XY}&&=\frac{1}{M_H^2}C^1_XC^2C^4_Y(C^3_{X'}m_{l_{\alpha}}{\cal C}_1-C^3_XM_3{\cal C}_0).
\end{eqnarray}
For $h$ and $\chi^0$ mediated diagram in Fig.\ref{h} (c), the couplings $C^1_X$ and $C^3_X$ are same with those in Eq.(\ref{E1}) except an interchange of subscripts $(i\leftrightarrow k,j\leftrightarrow k)$. The remaining coupling $C^2$ and masses denotation are
\begin{eqnarray}
C^2&&=\sum_{a=1,2,3}\frac{i}{4}\big(2(-2 v_dZ^{E\ast}_{i(3+a)}Y_{l_{a}}Y_{l_{a}}
Z^{E}_{j(3+a)}Z^h_{l1}-2v_dZ^{E\ast}_{ia}Y_{l_{a}}Y_{l_{a}}
Z^{E}_{ja}Z^h_{l1}\nonumber\\
&&+g_1Z^{E\ast}_{i(3+\beta)}Z^{E}_{j(3+\beta)}(g_1v_dZ^h_{l1}-g_1v_uZ^h_{l2}
-4M^B_DZ^h_{l3}))+Z^{E\ast}_{ia}Z^{E\ast}_{ja}\nonumber\\
&&\times(4(g_1M_D^BZ^h_{l3}+g_2M_D^WZ^h_{l4})
+(g_2^2-g_1^2)v_dZ^h_{l1}+(g_1^2-g_2^2)v_uZ^h_{l2})\big),\nonumber\\
M_1&&=m^i_{\tilde{e}},M_2=m^j_{\tilde{e}},M_3=m^k_{\chi^{0}}.\label{E6}
\end{eqnarray}
For $h$ and $\chi^{0c} $ mediated diagram in Fig.\ref{h} (c), the couplings $C^1_X$ and $C^3_X$ are same with those in Eq.(\ref{E3}) except an interchange of subscripts $(i\leftrightarrow k,j\leftrightarrow k)$. The couplings $C^2$ and masses denotation are same with that Eq.(\ref{E6}).For $A^0$ mediated diagrams in Fig.\ref{h} (c), the contribution is zero as we have assumed both $M_D^W$ and $M_D^B$ are real numbers in the coupling of $A^{0} \tilde{e}\tilde{e}$ interaction.

For $h$ and $\chi^{\pm}$ mediated diagram in Fig.\ref{h} (d), the couplings $C^1_X$ and $C^3_X$ are same with those in Eq.(\ref{E4}) except an interchange of subscripts $(i\leftrightarrow k,j\leftrightarrow k)$. The remaining coupling $C^2$ and mass denotation are
\begin{eqnarray}
C^2&&=\frac{i}{4}\delta^{ij}\big(4(g_1M_D^BZ^h_{l3}-g_2M_D^WZ^h_{l4})-(g_1^2+g_2^2)
(v_dZ^h_{l1}-v_uZ^h_{l2})\big),\nonumber\\
M_1&&=m^i_{\tilde{\nu}},M_2=m^j_{\tilde{\nu}},M_3=m^k_{\chi^{\pm}}.\label{E9}
\end{eqnarray}
For $A^0$ mediated diagrams in Fig.\ref{h} (d), the contribution is also zero since we have assumed both $M_D^W$ and $M_D^B$ are real numbers in the coupling of $A^{0} \tilde{\nu}\tilde{\nu}$ interaction.

%%%%%%%%%%%%%%%%%%%%%%%%%%%%%%%%%%%%%%%%%%%%%%%%%%%%%%%%%%%%%%%%%%%
\begin{figure}[htbp]
\setlength{\unitlength}{1mm}
\begin{minipage}[c]{1\textwidth}
\centering
\includegraphics[width=4.0in]{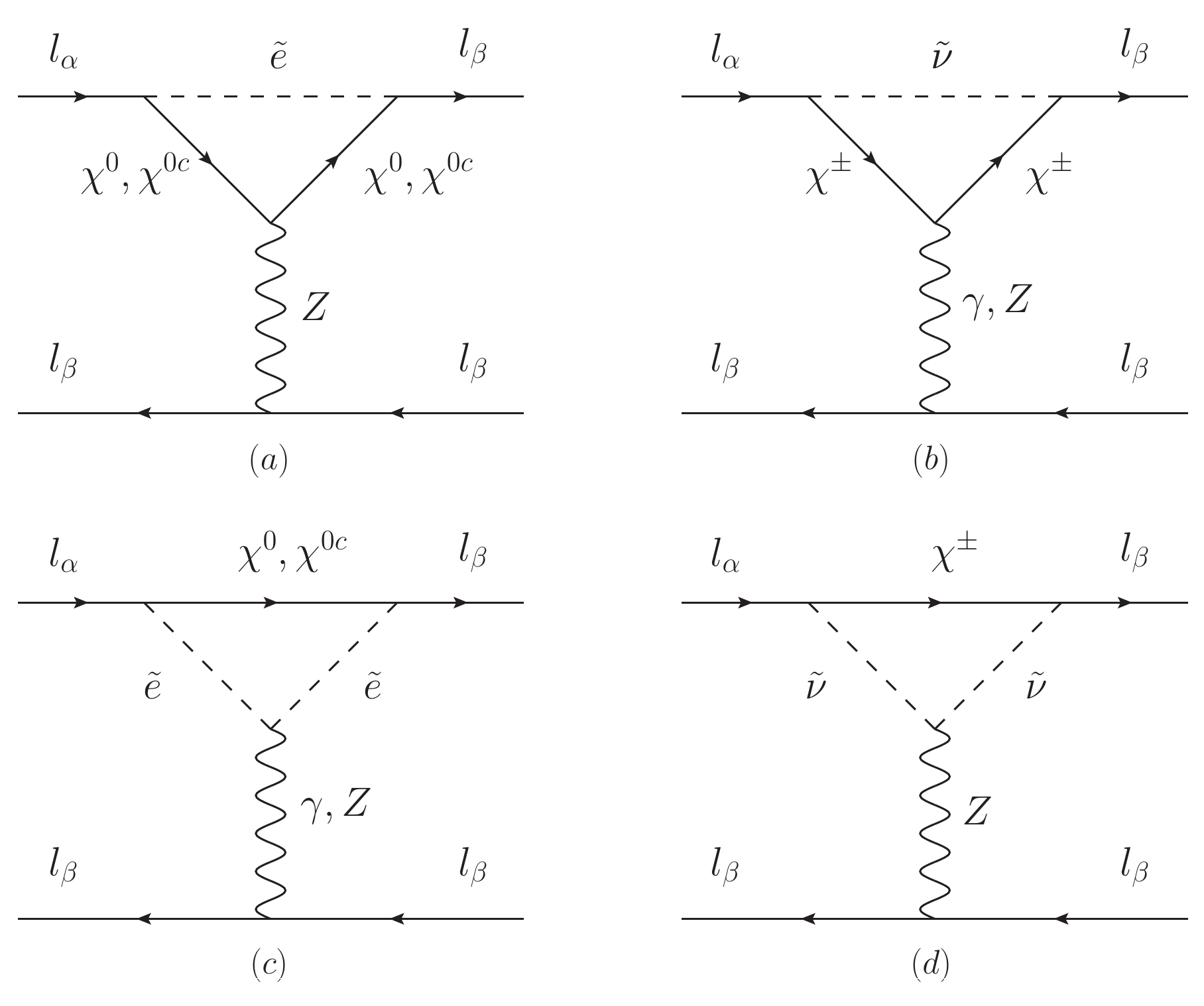}
\end{minipage}
\caption[]{Photon and Z penguin diagrams contributing to $l_{\alpha}\rightarrow 3l_{\beta}$ in MRSSM.}
\label{gz}
\end{figure}
%%%%%%%%%%%%%%%%%%%%%%%%%%%%%%%%%%%%%%%%%%%%%%%%%%%%%%%%%%%%%%%%%%%%%%%%%%%%%%%%%%

The photon and Z boson mediated diagrams contributing to $l_{\alpha}\rightarrow 3l_{\beta}$ in MRSSM are presented in Fig.\ref{gz}.
The coefficients in Fig.\ref{gz} (a,b) are calculated by
\begin{eqnarray}
A^V_{XY}&&=\frac{-1}{M_Z^2}C^1_{X'}C^4_Y\big(C^2_{X'}C^3_X {\cal B}_0(0,M_2,M_1)+(C^2_{X'}C^3_XM_3^2-C^2_XC^3_XM_1M_2\nonumber\\
&&-C^2_XC^3_{X'}M_1m_{l_{\alpha}}){\cal C}_0-C^2_XC^3_{X'}M_1m_{l_{\alpha}}{\cal C}_1-2C^2_{X'}C^3_X{\cal C}_{00}\nonumber\\
&&+C^2_{X'}C^3_Xm_{l_{\alpha}}^2{\cal C}_{1}+C^2_{X'}C^3_{X'}M_2m_{l_{\alpha}}{\cal C}_1\big),
\end{eqnarray}
where $C^4_X$ are identical in Fig.\ref{gz}(a-d),
\begin{eqnarray}
C^4_L=\frac{i}{2}(g_2 c_w-g_1 s_w),C^4_R=-ig_1 s_w.
\end{eqnarray}
For $\chi^0$ mediated diagram in Fig.\ref{gz} (a), the couplings $C^1_X$, $C^3_X$ and masses denotation are same with those in Eq.(\ref{E1}). The remaining couplings are
\begin{eqnarray}
C^2_L&&=\frac{i}{2}(g_1 s_w+g_2 c_w)(N^{1\ast}_{j3}N^{1}_{i3}-N^{1\ast}_{j4}N^{1}_{i4}),\nonumber\\
C^2_R&&=\frac{i}{2}(g_1 s_w+g_2 c_w)(N^{2\ast}_{i3}N^{2}_{j3}-N^{2\ast}_{i4}N^{2}_{j4}).\label{E7}
\end{eqnarray}
For $\chi^{0c}$ mediated diagram in Fig.\ref{gz} (a), the couplings $C^1_X$, $C^3_X$ and masses denotation are same with those in Eq.(\ref{E3}). The remaining couplings $C^2_X$ are same with those in Eq.(\ref{E7}).
For $\chi^{\pm}$ mediated diagram in Fig.\ref{gz} (b), the couplings $C^1_X$, $C^3_X$ and masses denotation are same with those in Eq.(\ref{E4}). The remaining couplings are
\begin{eqnarray}
C^2_L&&=\frac{-i}{2}\big(2g_2 c_w V^{1\ast}_{j1}V^{1}_{i1}+
(g_2 c_w-g_1 s_w)V^{1\ast}_{j2}V^{1}_{i1}\big),\nonumber\\
C^2_R&&=\frac{-i}{2}\big(2g_2 c_w U^{1\ast}_{i1}U^{1}_{j1}+
(g_2 c_w-g_1 s_w)U^{1\ast}_{i2}U^{1}_{j2}\big).
\end{eqnarray}

The coefficients in Fig.\ref{gz} (c,d) are calculated by
\begin{eqnarray}
A^V_{XY}&&=\frac{1}{M_Z^2}C^1_{X'}C^2C^3_XC^4_Y{\cal C}_{00}.
\end{eqnarray}
For $\chi^{0}$ mediated diagram in Fig.\ref{gz} (c), the couplings $C^1_X$ and $C^3_X$ are same with those in Eq.(\ref{E1}) except an interchange of subscripts $(i\leftrightarrow k,j\leftrightarrow k)$. The masses denotation are same with those in Eq.(\ref{E6}), and the remaining coupling $C^2$ is
\begin{eqnarray}
C^2&&=\sum_{a=1,2,3}\frac{i}{2}\big(-2g_1s_w Z^{E\ast}_{i(3+a)}Z^{E}_{j(3+a)}+(g_2 c_w-g_1s_w)Z^{E\ast}_{ia}Z^{E}_{ja}\big).\label{E8}
\end{eqnarray}
For $\chi^{0c}$ mediated diagram in Fig.\ref{gz} (c), the couplings $C^1_X$ and $C^3_X$ are same with those in Eq.(\ref{E3}) except an interchange of subscripts $(i\leftrightarrow k,j\leftrightarrow k)$. The masses denotation are same with those in Eq.(\ref{E6}). The remaining coupling $C^2$ is same with that in Eq.(\ref{E8}).
For $\chi^{\pm}$ mediated diagram in Fig.\ref{gz} (d), the couplings $C^1_X$ and $C^3_X$ are same with those in Eq.(\ref{E4}) except an interchange of subscripts $(i\leftrightarrow k,j\leftrightarrow k)$. The masses denotation are same with those in Eq.(\ref{E9}). The remaining coupling $C^2$ is
$C^2=-\frac{i}{2}\delta^{ij}(g_1s_w+g_2c_w)$.

The $K^X_1$ and $K^X_2$ coefficients in Fig.\ref{gz} (b) are calculated by
\begin{eqnarray}
K^X_1&&=\frac{I}{16\pi^2}C^1_{X'}C^2_{X'}C^3_{X}{\cal C}_{12},\nonumber\\
K^X_2&&=\frac{-I}{16\pi^2m_{l_{\alpha}}}C^1_{X}\big(C^2_{X}((C^3_{X}M_2+C^3_{X'}m_{l_{\alpha}}){\cal C}_1\nonumber\\
&&+C^3_{X'}m_{l_{\alpha}}({\cal C}_{12}+{\cal C}_{11}))+C^2_{X'}C^3_{X}M_1{\cal C}_{2}\big).
\end{eqnarray}
The couplings $C^1_X$, $C^3_X$ and masses denotation are same with those in Eq.(\ref{E4}), and $C^2_X=-ie\delta^{ij}$.

The coefficient $K^X_1$ in Fig.\ref{gz} (c) is zero, and $K^X_2$ is calculated by
\begin{eqnarray}
K^X_2&&=\frac{I}{32\pi^2m_{l_{\alpha}}}C^1_{X}C^2\big(C^3_{X'}m_{l_{\alpha}}(2{\cal C}_{12}+2{\cal C}_{11}\nonumber\\
&&+{\cal C}_{1})-C^3_{X}M_3({\cal C}_0+2{\cal C}_1+2{\cal C}_2)\big).
\end{eqnarray}
For $\chi^{0c}$ mediated diagram in Fig.\ref{gz} (c), the couplings $C^1_X$ and $C^3_X$ are same with those in Eq.(\ref{E3}) except an interchange of subscripts $(i\leftrightarrow k,j\leftrightarrow k)$. The masses denotation are same with those in Eq.(\ref{E6}). The remaining coupling $C^2$ is $ie\delta^{ij}$.
For $\chi^{\pm}$ mediated diagram in Fig.\ref{gz} (d), the couplings $C^1_X$ and $C^3_X$ are same with those in Eq.(\ref{E4}) except an interchange of subscripts $(i\leftrightarrow k,j\leftrightarrow k)$. The masses denotation are same with those in Eq.(\ref{E9}). The remaining coupling $C^2$ is $ie\delta^{ij}$.

%%%%%%%%%%%%%%%%%%%%%%%%%%%%%%%%%%%%%%%%%%%%%%%%%%%%
\begin{figure}[htbp]
\setlength{\unitlength}{1mm}
\begin{minipage}[c]{1\textwidth}
\centering
\includegraphics[width=4.0in]{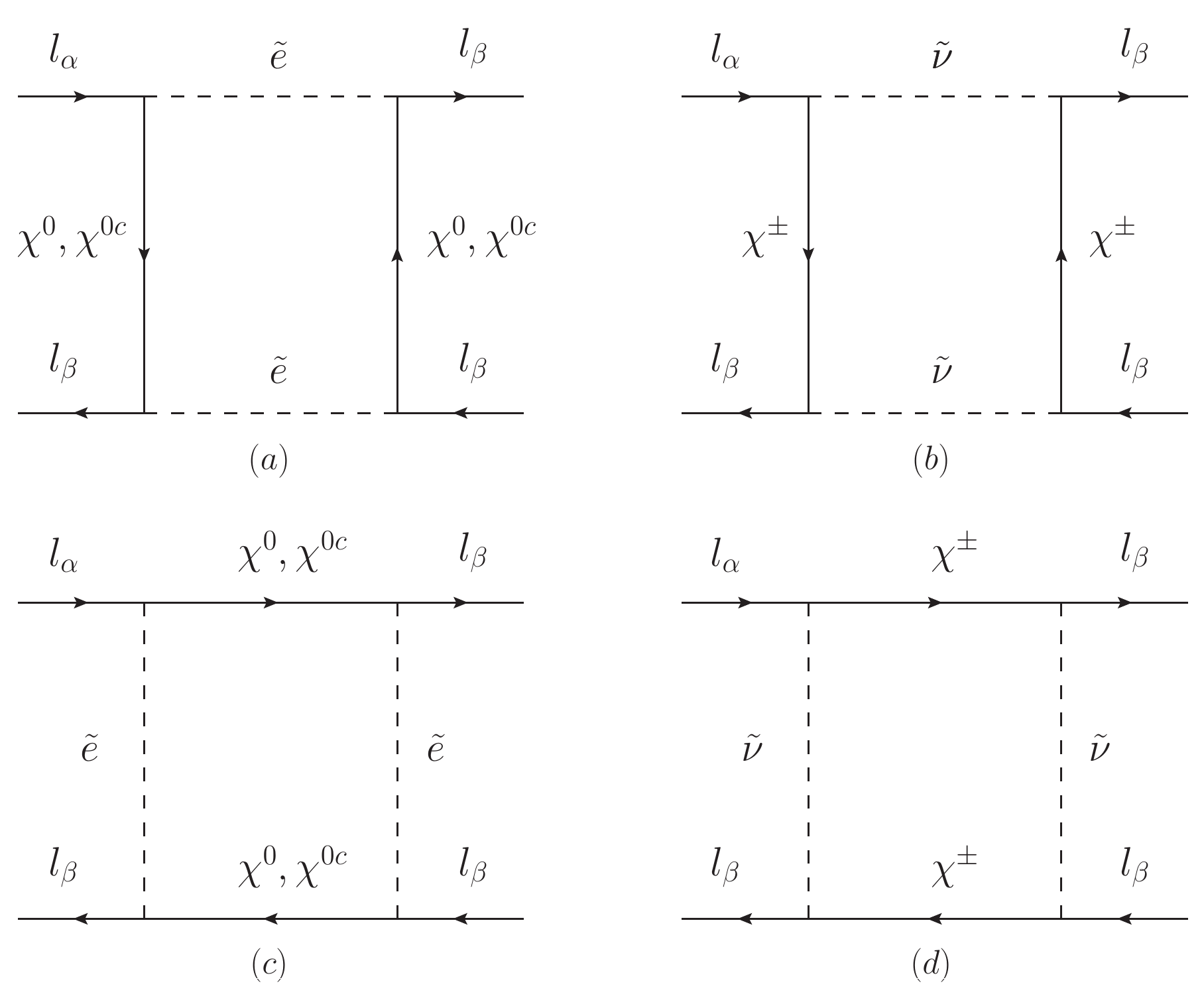}
\end{minipage}
\caption[]{Box diagrams contributing to $l_{\alpha}\rightarrow 3l_{\beta}$ in MRSSM.}
\label{b}
\end{figure}
%%%%%%%%%%%%%%%%%%%%%%%%%%%%%%%%%%%%%%%%%%%%%%%%%%%%%%%%%%%%%%%%%%%
The box diagrams contributing to $l_{\alpha}\rightarrow 3l_{\beta}$ in MRSSM are presented in Fig.\ref{gz}.
The coefficients in Fig.\ref{b} (a,b) are calculated by
\begin{eqnarray}
A^S_{XY}&&=C^1_{X}C^2_{X}C^3_{Y}M_1\big((C^4_YM_3-C^4_{Y'}m_{l_{\alpha}}){\cal D}_0-C^4_{Y'}m_{l_{\alpha}}({\cal D}_2+{\cal D}_1)\big),\nonumber\\
A^V_{XY}&&=C^1_{X'}C^2_XC^3_{Y'}C^4_Y{\cal D}_{00}.
\end{eqnarray}
For two $\chi^0$ mediated diagram in Fig.\ref{b} (a), the couplings are
\begin{eqnarray}
C^1_L&&=-i\sqrt{2}g_1N^{1\ast}_{i1}Z^{E\ast}_{k(3+\beta)},
C^1_R=-iY_{l_{\beta}}Z^{E}_{k(3+\beta)}N^2_{i3},\nonumber\\
C^2_L&&=-iN^{2\ast}_{i3}Y_{l_{\beta}}Z^{E}_{l(3+\beta)},
C^2_R=-i\sqrt{2}g_1Z^{E}_{l(3+\beta)}N^1_{i1}.
\label{E10}
\end{eqnarray}
The couplings $C^3_X$ are same with $C^1_X$ in Eq.(\ref{E10}) with interchange of subscripts $(i\leftrightarrow j,k\leftrightarrow l)$, $C^4_X$ are same with $C^2_X$ in Eq.(\ref{E10}) with index exchange $(i\leftrightarrow j,k\leftrightarrow l,\beta\leftrightarrow \alpha)$.
For two $\chi^{0c}$ mediated diagram in Fig.\ref{b} (a), the couplings are
\begin{eqnarray}
C^1_L&&=-iY_{l_{\beta}}N^{2\ast}_{i3}Z^{E\ast}_{k\beta},
C^1_R=\frac{i}{\sqrt{2}}Z^{E\ast}_{k\beta}(g_1N^1_{i1}+g_2N^1_{i2}),\nonumber\\
C^2_L&&=\frac{i}{\sqrt{2}}Z^{E}_{l\beta}(g_1N^{1\ast}_{i1}+g_2N^{1\ast}_{i2}),
C^2_R=-iY_{l_{\beta}}Z^{E}_{l\beta}N^2_{i3}.
\label{E11}
\end{eqnarray}
The couplings $C^3_X$ are same with $C^1_X$ in Eq.(\ref{E11}) with interchange of subscripts $(i\leftrightarrow j,k\leftrightarrow l)$, $C^4_X$ are same with $C^2_X$ in Eq.(\ref{E11}) with index exchange $(i\leftrightarrow j,k\leftrightarrow l,\beta\leftrightarrow \alpha)$.
For $\chi^{0}\chi^{0c}$ mediated diagram in Fig.\ref{b} (a), the couplings $C^1_X$ and $C^2_X$ are same with those in Eq.(\ref{E11}), and the couplings $C^3_X$ and $C^4_X$ are same with those after Eq.(\ref{E10}). For $\chi^{0c}\chi^{0}$ mediated diagram in Fig.\ref{b} (a), the couplings $C^1_X$ and $C^2_X$ are same with those in Eq.(\ref{E10}), and the couplings $C^3_X$ and $C^4_X$ are same with those after Eq.(\ref{E11}). The masses denotation are $M_1=m^i_{\chi^0}$, $M_2=m^l_{\tilde{e}}$, $M_3=m^j_{\chi^0}$ and $M_4=m^k_{\tilde{e}}$.

For two $\chi^{\pm}$ mediated diagram in Fig.\ref{b} (b), the couplings and masses denotation are
\begin{eqnarray}
C^1_L&&=iU^{1\ast}_{i2}Z^V_{k\beta}Y_{l_{\beta}},
C^1_R=-ig_2Z^{V\ast}_{k\beta}V^1_{i1},
C^2_L=-(C^1_R)^{\ast}(k\leftrightarrow l),\nonumber\\
C^2_R&&=-(C^1_L)^{\ast}(k\leftrightarrow l),
C^3_L=-C^1_L(i\leftrightarrow j,k\leftrightarrow l),
C^3_R=-C^1_R(i\leftrightarrow j,k\leftrightarrow l),\nonumber\\
C^4_L&&=-(C^1_R)^{\ast}(i\leftrightarrow j,k\leftrightarrow l,\beta\leftrightarrow \alpha),
C^4_R=-(C^1_L)^{\ast}(i\leftrightarrow j,k\leftrightarrow l,\beta\leftrightarrow \alpha).
\end{eqnarray}
The masses denotation are $M_1=m^i_{\chi^{\pm}}$, $M_2=m^l_{\tilde{\nu}}$, $M_3=m^j_{\chi^{\pm}}$ and $M_4=m^k_{\tilde{\nu}}$.

The coefficients in Fig.\ref{b} (c,d) are calculated by
\begin{eqnarray}
A^S_{XY}&&=C^1_{X}C^2_{Y}C^3_{Y}M_2\big(C^4_XM_4{\cal D}_0-C^4_{X'}m_{l_{\alpha}}({\cal D}_2+{\cal D}_1)\big),\nonumber\\
A^V_{XY}&&=C^1_{X'}C^2_YC^3_{Y'}C^4_X{\cal D}_{00}.
\end{eqnarray}
The couplings $C^1_X$, $C^2_X$, $C^3_X$ and $C^4_X$ correspond to diagrams in Fig.\ref{b} (c,d) are same as those in Fig.\ref{b} (a,b) respectively, where following  interchanges of subscripts should be made: $(i\leftrightarrow k)$, $(i\leftrightarrow l)$, $(j\leftrightarrow l)$ and $(j\leftrightarrow k)$. The masses notation in Fig.\ref{b} (c) are $M_1=m^i_{\tilde{e}}$, $M_2=m^l_{\chi^0}$, $M_3=m^j_{\tilde{e}}$ and $M_4=m^k_{\chi^0}$. The masses notation in Fig.\ref{b} (d) are $M_1=m^i_{\tilde{\nu}}$, $M_2=m^l_{\chi^{\pm}}$, $M_3=m^j_{\tilde{\nu}}$ and $M_4=m^k_{\chi^{\pm}}$.

Using the Wilson coefficients in Eqs.(\ref{llg}, \ref{4l}), the decay width $\Gamma(l_{\alpha}\rightarrow 3 l_{\beta})$ is given by \cite{Flavor}
\begin{eqnarray}
\Gamma(l_{\alpha}\rightarrow 3 l_{\beta})&&=\frac{m^5_{l_{\alpha}}}{512\pi^3}[e^4(|K^L_2|^2+|K^R_2|^2)(\frac{16}{3}ln\frac{m_{l_1}}{m_{l_2}}-\frac{22}{3})+\frac{1}{24}(|A^S_{LL}|^2+|A^S_{RR}|^2)\nonumber\\
&&+\frac{1}{12}(|A^S_{LR}|^2+|A^S_{RL}|^2)+\frac{2}{3}(|\hat{A}^V_{LL}|^2+|\hat{A}^V_{RR}|^2)+\frac{1}{3}(|\hat{A}^V_{LR}|^2+|\hat{A}^V_{RL}|^2)\nonumber\\
&&+6(|A^T_{LL}|^2+A^T_{RR}|^2)+\frac{2e^2}{3}Re(K^L_2A^{S\ast}_{RL}+K^R_2A^{S\ast}_{LR})-\frac{4e^2}{3}Re(K^L_2\hat{A}^{V\ast}_{RL}\nonumber\\
&&+K^R_2\hat{A}^{V\ast}_{LR})-\frac{8e^2}{3}Re(K^L_2\hat{A}^{V\ast}_{RR}+K^R_2\hat{A}^{V\ast}_{LL})-Re(A^S_{LL}A^{T\ast}_{LL}+A^S_{RR}A^{T\ast}_{RR})\nonumber\\
&&-\frac{1}{3}Re(A^S_{LR}\hat{A}^{V\ast}_{LR}+A^S_{RL}\hat{A}^{T\ast}_{RL})].
\end{eqnarray}

As mentioned earlier, loop integrals are given in term of Passarino-Veltman functions\cite{PVI},
\begin{eqnarray}
{\cal C}_{(0,1,...,12)}&=&\frac{i}{16\pi^2}{\cal C}_{(0,1,...,12)}
(m^2_{l_{\alpha}},0,0;M_3,M_1,M_2),\nonumber\\
{\cal D}_{(0,1,...,00)}&=&\frac{i}{16\pi^2}{\cal D}_{(0,1,...,00)}
(0,0,m^2_{l_{\alpha}},0,;m^2_{l_{\alpha}},0;M_1,M_2,M_3,M_4).
\end{eqnarray}
The explicit expressions of these loop integrals are given in Refs \cite{collier1,collier2,collier3} and $\overline{MS}$ scheme is used to delete the infinite terms. These loop integrals can be calculated through the Mathematica package Package-X \cite{X} and a link to Collier which is a fortran library for the numerical evaluation of one-loop scalar and tensor integrals\cite{collier}.

\section{Numerical Analysis}
\label{sec3}
In the numerical analysis, we will use the benchmark point in Refs.\cite{Die3,sks1,sks2} as the default values in our parameter setup, where the soft breaking terms $m_{l}^{2}$, $m_{r}^{2}$ are diagonal. In this work, the off-diagonal entries of the soft breaking terms $m_{l}^{2}$, $m_{r}^{2}$ are parameterized by mass insertion as in Ref.\cite{Rosiek3,Kss,Hzz},
\begin{eqnarray}
\Big(m^{2}_{l}\Big)^{IJ}&=&\delta ^{IJ}_{l}\sqrt{(m^{2}_{l})^{II}(m^{2}_{l})^{JJ}},\nonumber\\
\Big(m^{2}_{r}\Big)^{IJ}&=&\delta ^{IJ}_{r}\sqrt{(m^{2}_{r})^{II}(m^{2}_{r})^{JJ}},
\end{eqnarray}
where I,J=\{1,2,3\}. We also assume $\delta ^{IJ}_{l}$ = $\delta ^{IJ}_{r}$ = $\delta ^{IJ}$. In the following, we will use LFV decays $l_\alpha\rightarrow l_\beta\gamma$ to constrain the parameters $\delta ^{IJ}$ and the explicit expression can be found in Ref.\cite{sks2}.
For the values of $\mu_u(\mu_d)$, $M_D^W$ and $M_D^B$, we have considered the constraints from theoretical valid regions in Ref. \cite{PhD} and the experimental bounds from ATLAS\cite{AT1806}. The large value of $|v_T|$ is excluded by measurement of $W$ mass cause the vev $v_T$ of the $SU(2)_L$ triplet field $T^0$ gives a correction to $W$ mass through\cite{Die1}
\begin{eqnarray}
m_W^2=\frac{1}{4}g_2^2v^2+g_2^2v_T^2,
\label{}
\end{eqnarray}
with $v^2=v_u^2+v_d^2$.
Then, the numerical values in our parameter setup are
\begin{eqnarray}
&&\alpha_{em}(m_Z)=1/137, m_Z=91.1876 \text{ GeV}, m_W=80.379 \text{ GeV}, \nonumber\\
&&sin^2\theta_W=0.23129,m_e=0.510 \text{ MeV}, m_{\mu}=105.6 \text{ MeV}, m_{\tau}=1.776 \text{ GeV},\nonumber\\
&&tan\beta=40,B_{\mu}=300^2 \text{ GeV}^2,\lambda_d=-\lambda_u=0.15,\Lambda_d=-1.0,\Lambda_u=-1.15, \nonumber\\
&&v_S=-0.14\text{ GeV},v_T=-0.34\text{ GeV},M_D^B=M_D^W=\mu_d=\mu_u=600 \text{ GeV},\nonumber\\
&&m_T^2=3000^2 \text{ GeV}^2,m^2_l=m_r^2=1000^2 \text{ GeV}^2.
\label{values}
\end{eqnarray}

%%%%%%%%%%%%%%%%%%%%%%%%%%%%%%%%%%%%%%%%%%%%%%%%%%%%%%%%%%%%%%%%%%%
\begin{figure*}
\centering
\includegraphics[width=4.0in]{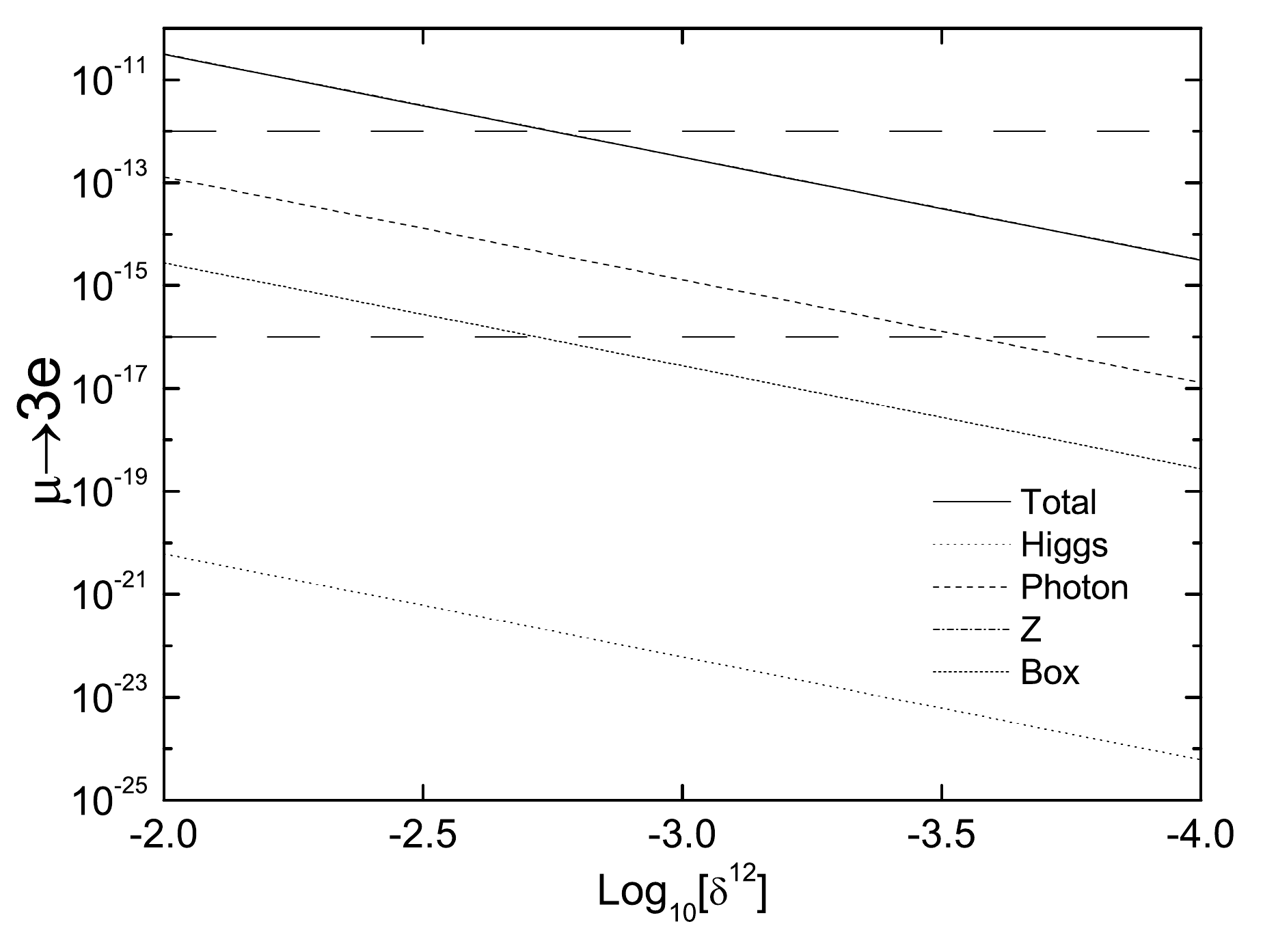}
\caption[]{Br$(\mu\rightarrow 3 e)$ vary as a function of $\text{Log}_{10}[\delta^{12}]$ in MRSSM, where the contributions from total diagrams (solid line), Higgs penguins (dot line), $\gamma$ penguins (dash line), Z penguins (dash dot line) and box diagrams (short dash line) are listed. The upper horizontal dash line denotes the experimental upper limit and the lower horizontal dash line denotes the future experimental sensitivity.}
\label{fig1}
\end{figure*}
%%%%%%%%%%%%%%%%%%%%%%%%%%%%%%%%%%%%%%%%%%%%%%%%%%%%%%%%%%%%%%%%%%%

Taking data in Eq.(\ref{values}) and $\delta^{13}$ = $\delta^{23}$ =0, we display the theoretical prediction of Br$(\mu\rightarrow 3 e)$ versus $\text{Log}_{10}[\delta^{12}]$ in MRSSM in Fig.\ref{fig1}, where the contributions from total diagrams (solid line), Higgs penguins (dot line), $\gamma$ penguins (dash line), Z penguins (dash dot line) and box diagrams (short dash line) are listed.
We observe that a linear relationship is displayed between different predictions for Br$(\mu\rightarrow 3 e)$ and $\text{Log}_{10}[\delta^{12}]$ in logarithmic scale, which show a great dependence of Br$(\mu\rightarrow 3 e)$ on $\delta^{12}$. It shows that Higgs contribution is negligible (${\cal O}(10^{-25}-10^{-21})$), which is ten orders of magnitude below the total prediction for Br$(\mu\rightarrow 3 e)$. The box contribution (${\cal O}(10^{-19}-10^{-15})$) and $\gamma$ contribution (${\cal O}(10^{-16}-10^{-13})$) are about four and two orders of magnitude below the total prediction respectively. The contribution from Z diagrams takes an important role in prediction for Br$(\mu\rightarrow 3 e)$ and is too close to the total prediction to distinguish them in Fig.\ref{fig1}. Considering the discussion in Ref.\cite{sks2}, the value of $\delta^{13}$ is about $10^{-1}$. Then, the total prediction for Br$(\mu\rightarrow 3 e)$(${\cal O}(10^{-9})$)  is one order of magnitude below the current experimental limit in Table.\ref{Tab1}.

%%%%%%%%%%%%%%%%%%%%%%%%%%%%%%%%%%%%%%%%%%%%%%%%%%%%%%%%%%%%%%%%%%%
\begin{figure*}
\centering
\includegraphics[width=4.0in]{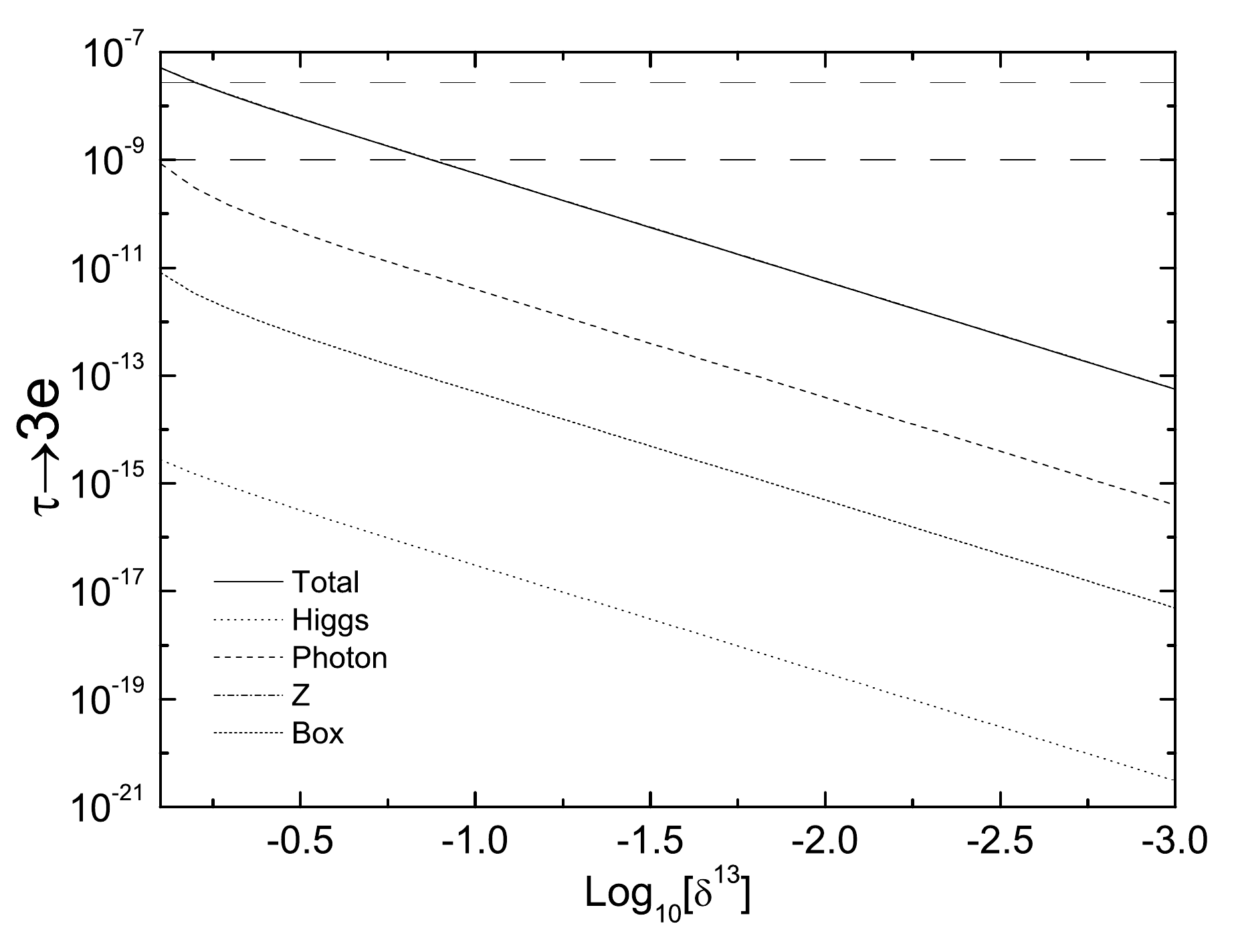}
\caption[]{$Br(\tau\rightarrow 3 e)$ vary as a function of $\text{Log}_{10}[\delta^{13}]$ in MRSSM, where the contributions from total diagrams (solid line), Higgs penguins (dot line), $\gamma$ penguins (dash line), Z penguins (dash dot line) and box diagrams (short dash line) are listed. The upper horizontal dash line denotes the experimental upper limit and the lower horizontal dash line denotes the future experimental sensitivity.}
\label{fig2}
\end{figure*}
%%%%%%%%%%%%%%%%%%%%%%%%%%%%%%%%%%%%%%%%%%%%%%%%%%%%%%%%%%%%%%%%%%%
Taking data in Eq.(\ref{values}) and $\delta^{12}$ = $\delta^{23}$ =0, we display the theoretical prediction of Br$(\tau\rightarrow 3 e)$ versus $\text{Log}_{10}[\delta^{13}]$ in MRSSM in Fig.\ref{fig2}, where the contributions from total diagrams (solid line), Higgs penguins (dot line), $\gamma$ penguins (dash line), Z penguins (dash dot line) and box diagrams (short dash line) are listed.
We observe that a linear relationship is displayed between different prediction for Br$(\tau\rightarrow 3 e)$ and $\text{Log}_{10}[\delta^{12}]$ in logarithmic scale, which shows the great dependence of Br$(\tau\rightarrow 3 e)$ on $\delta^{12}$. It shows that Higgs contribution is negligible (${\cal O}(10^{-21}-10^{-15})$), which is eight orders of magnitude below the total prediction. The box contribution (${\cal O}(10^{-18}-10^{-11})$) and $\gamma$ contribution (${\cal O}(10^{-15}-10^{-9})$) are about four and two orders of magnitude below the total prediction respectively. The contribution from Z diagrams is very close to the total prediction and takes an important role in Br$(\tau\rightarrow 3 e)$, which is hard to distinguish them in Fig.\ref{fig2}. Considering the discussion in Ref.\cite{sks2}, the value of $\delta^{13}$ is about $10^{-3}$. Then, the total prediction Br$(\tau\rightarrow 3 e)$(${\cal O}(10^{-9})$) is one order of magnitude below the current experimental limit in Table.\ref{Tab1}.

%%%%%%%%%%%%%%%%%%%%%%%%%%%%%%%%%%%%%%%%%%%%%%%%%%%%%%%%%%%%%%%%%%%
\begin{figure*}
\centering
\includegraphics[width=4.0in]{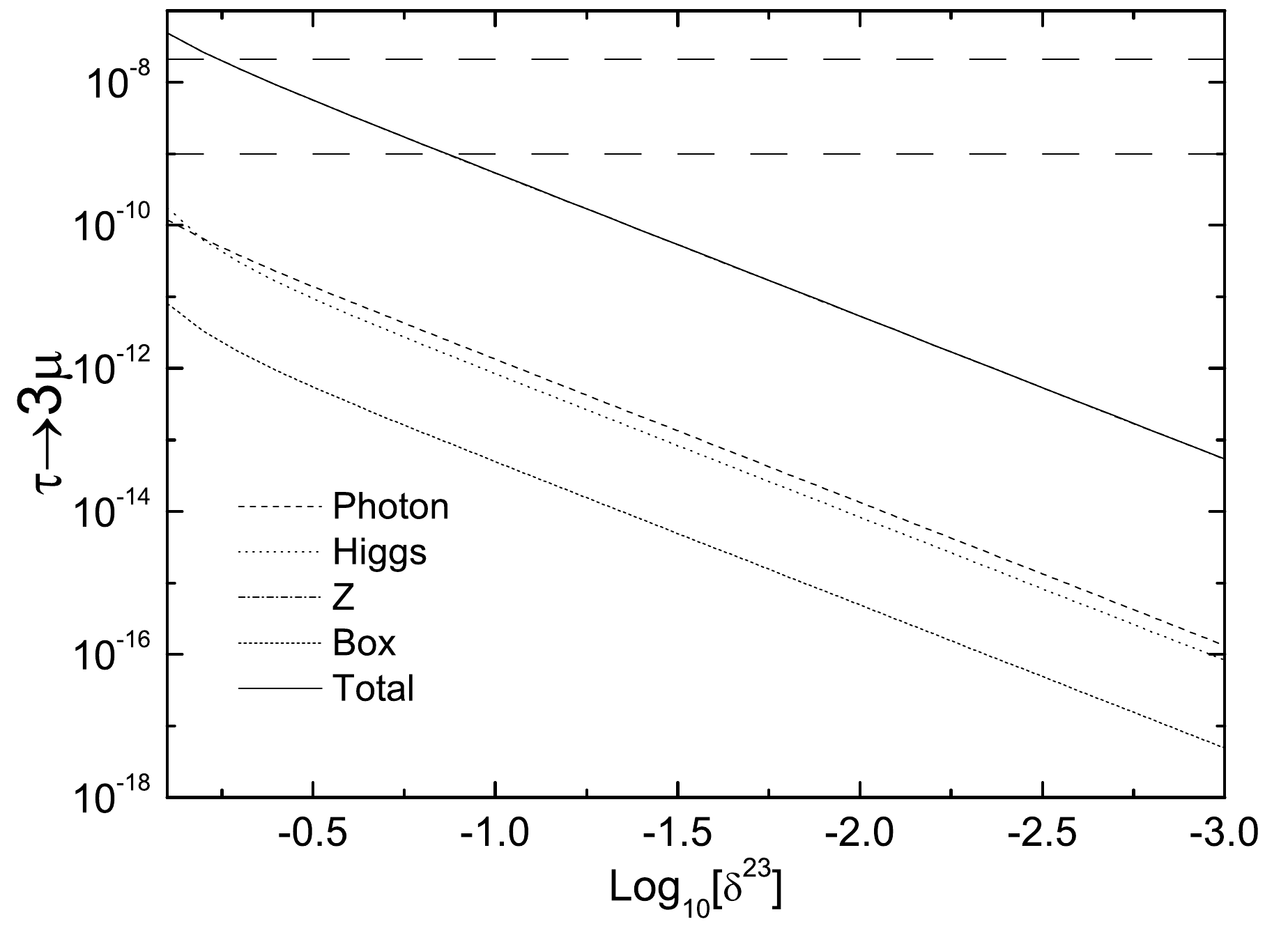}
\caption[]{$Br(\tau\rightarrow 3 \mu)$ vary as a function of $\text{Log}_{10}[\delta^{23}]$ in MRSSM,where the contributions from total diagrams (solid line), Higgs penguins (dot line), $\gamma$ penguins (dash line), Z penguins (dash dot line) and box diagrams (short dash line) are listed. The upper horizontal dash line denotes the experimental upper limit and the lower horizontal dash line denotes the future experimental sensitivity.}
\label{fig3}
\end{figure*}
%%%%%%%%%%%%%%%%%%%%%%%%%%%%%%%%%%%%%%%%%%%%%%%%%%%%%%%%%%%%%%%%%%
Taking data in Eq.(\ref{values}) and $\delta^{12}$ = $\delta^{13}$ =0, we display the theoretical prediction of Br$(\tau\rightarrow 3 \mu)$ versus $\text{Log}_{10}[\delta^{23}]$ in MRSSM in Fig.\ref{fig3}, where the contributions from total diagrams (solid line), Higgs penguins (dot line), $\gamma$ penguins (dash line), Z penguins (dash dot line) and box diagrams (short dash line) are listed.
There is also a linear relationship between different prediction for Br$(\tau\rightarrow 3 \mu)$ and $\text{Log}_{10}[\delta^{23}]$ in logarithmic scale, which shows the great dependence of Br$(\tau\rightarrow 3 \mu)$ on $\delta^{23}$. Compare with other three contributions, it shows that box contribution is negligible (${\cal O}(10^{-18}-10^{-12})$). The Higgs contribution (${\cal O}(10^{-17}-10^{-10})$) and $\gamma$ contribution (${\cal O}(10^{-16}-10^{-10})$) are about two orders of magnitude below the total prediction respectively. The contribution from Z penguins is very close to the total prediction and takes an important role in Br$(\tau\rightarrow 3 \mu)$, which is hard to distinguish them in Fig.\ref{fig3}. Considering the discussion in Ref.\cite{sks2}, the value of $\delta^{23}$ is about $10^{-3}$. Then, the total prediction for Br$(\tau\rightarrow 3 \mu)$(${\cal O}(10^{-10})$)  is two orders of magnitude below the current experimental limit in Table.\ref{Tab1}.

\section{Conclusions}
\label{sec4}
We have investigated the LFV processes $l_{\alpha}\rightarrow 3l_{\beta}$ in the framework of Minimal R-symmetric
Supersymmetric Standard Model (MRSSM) as a function of model parameters $\delta^{ij}$. The predictions for Br$(l_{\alpha}\rightarrow 3l_{\beta})$ show a great dependent on off-diagonal inputs $\delta^{ij}$. Taking account of the constraints on $\delta^{ij}$ from LFV processes $l_{\alpha}\rightarrow l_{\beta}\gamma$, all predictions for Br$(l_{\alpha}\rightarrow 3l_{\beta})$ can be enhanced up to the current experimental limits or future experimental sensitivities. Thus, more precise measurements of Br$(l_{\alpha}\rightarrow l_{\beta}\gamma)$ and Br$(l_{\alpha}\rightarrow 3l_{\beta})$ in experiment are in need.

\section*{Acknowledgements}
The work has been supported by the Scientific Research Foundation of the Higher Education Institutions of Hebei Province with Grant No.BJ2019210, the Foundation of Baoding University with Grant No.2018Z01, the National Natural Science Foundation of China (NNSFC) with Grants No.11805140 and No.11705045, the Scientific and Technological Innovation Programs of Higher Education Institutions in Shanxi with Grant No.2017113, the Natural Science Foundation of Shanxi Province with Grant No. 201801D221021, the youth top-notch talent support program of the Hebei Province.


\begin{thebibliography}{0}
\bibitem{Kribs}
G. D. Kribs, E. Poppitz and N. Weiner, Phys. Rev. D 78 (2008) 055010.
\bibitem{Fayet}
P. Fayet, Nucl. Phys. B 90 (1975) 104.
\bibitem{Salam}
A. Salam, J. Strathdee, Nucl. Phys. B 87 (1975) 85.
\bibitem{Die1}
P. Diessner, W. Kotlarski, PoS CORFU 2014 (2015) 079.
\bibitem{Die2}
P. Diessner, J. Kalinowski, W. Kotlarski, D. St\"{o}ckinger, Adv. High Energy Phys. 2015 (2015) 760729.
\bibitem{Die3}
P. Diessner, J. Kalinowski, W. Kotlarski, D. St\"{o}ckinger, JHEP 1412 (2014) 124.
\bibitem{Die4}
P. Diessner, W. Kotlarski, S. Liebschner, D. St\"{o}ckinger, JHEP 1710 (2017) 142.
\bibitem{Die5}
P. Diessner, J. Kalinowski, W. Kotlarski, D. St\"{o}ckinger, JHEP 1603 (2016) 007.
\bibitem{Die6}
P. Diessner, G. Weiglein, JHEP 1907 (2019) 011.
\bibitem{Kumar}
A. Kumar, D. Tucker-Smith, N. Weiner, JHEP 1009 (2010) 111.
\bibitem{Blechman}
A. E. Blechman, Mod.Phys.Lett. A24 (2009) 633.
\bibitem{Kribs1}
G. D. Kribs, A. Martin, T. S. Roy, JHEP 0906 (2009) 042.
\bibitem{Frugiuele}
C. Frugiuele, T. Gregoire, Phys.Rev. D85 (2012) 015016.
\bibitem{Jan}
J. Kalinowski, Acta Phys.Polon. B47 (2016) 203.
\bibitem{Chakraborty}
S. Chakraborty, A. Chakraborty, S. Raychaudhuri, Phys.Rev. D 94 (2016) 035014.
\bibitem{Braathen}
J. Braathen, M. D. Goodsell, P. Slavich, JHEP 1609 (2016) 045.
\bibitem{Athron}
P. Athron, J.-hyeon Park, T. Steudtner, D. St\"{o}ckinger, A. Voigt, JHEP 1701 (2017) 079.
\bibitem{Alvarado}
C. Alvarado, A. Delgado, A. Martin, Phys. Rev. D97 (2018) 115044.

\bibitem{sks1}
K.-S. Sun, J.-B. Chen, X.-Y. Yang, H.-B. Zhang, Mod. Phys. Lett. A 34 (2019) 1950058.
\bibitem{sks2}
K.-S. Sun, J.-B. Chen, X.-Y. Yang, S.-K. Cui, Chin. Phys. C 43 (2019) 043101.
\bibitem{kss}
W. Kotlarski, D. St\"{o}ckinger, H. St\"{o}ckinger-Kim, arXiv:1902.06650.

\bibitem{tb1}
U. Bellgardt et al. [SINDRUM Collaboration], Nucl. Phys. B 299 (1988) 1.
\bibitem{tb2}
A. Blondel et al., arXiv:1301.6113 [physics.ins-det].
\bibitem{tb4}
K. Hayasaka et al., Phys. Lett. B 687, 139(2010).
\bibitem{tb3}
K. Hayasaka [Belle and Belle-II Collaborations], J. Phys. Conf. Ser. 408 (2013) 012069.
\bibitem{Pontecorvo1}
B.Pontecorvo, Zh. Eksp. Teor. Fiz. JETP 33 (1957) 549.

\bibitem{Pontecorvo2}
B.Pontecorvo, Zh. Eksp. Teor. Fiz. JETP 34(1958) 247.

\bibitem{Maki}
Z.Maki, M.Nakagawa and S.Sakata, Prog. Theor. Phys. 28 (1962) 870.


\bibitem{Flavor}
W. Porod, F. Staub, A. Vicente, Eur.Phys.J. C 74 (2014) 2992.
\bibitem{PVI}
G. Passarino, M.J.G. Veltman, Nucl. Phys. B 160 (1979) 151.
\bibitem{collier1}
A. Denner, S. Dittmaier, Nucl. Phys. B 658 (2003) 175.
\bibitem{collier2}
A. Denner, S. Dittmaier, Nucl. Phys. B 734 (2006) 62.
\bibitem{collier3}
A. Denner, Fortsch.Phys. 41 (1993) 307.
\bibitem{X}
H. H. Patel, Comput. Phys. Commun. 197 (2015) 276.
\bibitem{collier}
A. Denner, S. Dittmaier, L. Hofer, Comput.Phys.Commun. 212 (2017) 220.
\bibitem{Rosiek3}
J. Rosiek, P. Chankowski, A. Dedes, S. Jager, P. Tanedo, Comput.Phys.Commun. 181 (2010) 2180.

\bibitem{Kss}
K.-S. Sun, T.-F. Feng, T.-J. Gao, S.-M. Zhao, Nucl.Phys. B 865 (2012) 486.

\bibitem{Hzz}
H. B. Zhang, T. F. Feng, S. M. Zhao and F. Sun, Int.J.Mod.Phys. A 29 (2014) 1450123.

\bibitem{PhD}
P. Diessner, PhD thesis, Dresden, Tech. U., 2016.
\bibitem{AT1806}
ATLAS collaboration. (M. Aaboud et al.), Phys.Rev. D 98 (2018) 092012.






\end{thebibliography}
\end{document}